\documentclass[twocolumn]{scrartcl}

\usepackage[utf8]{inputenc}

\usepackage{fixltx2e}

\usepackage{amsthm}

\newtheoremstyle{named}{}{}{\itshape}{}{\bfseries}{.}{.5em}{\thmnote{#1 }#3}
\theoremstyle{named}
\newtheorem*{step}{Step}

\usepackage{microtype}
\usepackage{amsmath}
\usepackage{mathtools}
\usepackage{booktabs}
\usepackage{tabularx}

\usepackage{pgfplots}
\pgfplotsset{compat=newest}

\usepackage{siunitx}

\newcommand\mytitle{Algorithmic improvements for the CIECAM02 and CAM16 color appearance models}
\newcommand\myauthor{Nico Schlömer}

\usepackage[
  pdfencoding=unicode,
  ]{hyperref}
\hypersetup{
  pdfauthor={\myauthor},
  pdftitle={\mytitle}
}

\DeclarePairedDelimiter\abs{\lvert}{\rvert}%

\usepackage[T1]{fontenc}
\usepackage{newtxtext}
\usepackage{newtxmath}

\usepackage{gensymb}

\usepackage{amsmath}
\DeclareMathOperator{\sign}{sign}

\usepackage{bm}
\newcommand\rgb{\bm{R}}

\title{\mytitle\footnote{The LaTeX sources of this article are on \url{https://github.com/nschloe/note-on-cam16}}}
\author{\myauthor}

\begin{document}

\maketitle
\begin{abstract}
  This note is concerned with the CIECAM02 color appearance model and its
  successor, the CAM16 color appearance model. Several algorithmic flaws are
  pointed out and remedies are suggested. The resulting color model is
  algebraically equivalent to CIECAM02/CAM16, but shorter, more efficient, and
  works correctly for all edge cases.
\end{abstract}

\section{Introduction}

The CIECAM02 color appearance model~\cite{ciecam02} has attracted much
attention and was generally thought of as a successor to the ever so popular
CIELAB color model. However, it was quickly discovered that CIECAM02 breaks
down for certain input values. A fair number of research articles suggests
fixes for this behavior, most of them by modifying the first steps of the
forward model. Luo and Li give an overview of the suggested
improvements~\cite{ciecam02-recent}; see references therein.  Most recently,
Li and Luo~\cite{cam16} gave their own suggestion on how to best
circumvent the breakdown~\cite{cam16}. The updated algorithm differs from the
original CIECAM02 only in the first steps of the forward model.

It appears that that the rest of the algorithm has not received much attention
over the years. In both CIECAM02 and its updated version CAM16, some of the
steps are more complicated than necessary, and in edge cases lead to break
downs once again. The present document describes those flaws and suggests
improvements (Section~\ref{sec:ff}). The resulting model description
(Section~\ref{sec:full}) is entirely equivalent to the CIECAM02/CAM16, but is
simpler -- hence faster and easier to implement -- and works in all edge
cases.

All findings in this article are implemented in the open-source software package
colorio~\cite{colorio}.

\section{Flaws and fixes}\label{sec:ff}

This section describes the flaws of CAM16 and suggests fixes for them. Some of
them are trivial, others are harder to see. All listed steps also appear in the
CIECAM02 color appearance model and trivially apply there.

\subsection{Step 3, forward model}

The original Step 3 of the forward model reads

\begin{step}[3]
Calculate the postadaptation cone response
(resulting in dynamic range compression).
\[
  R_a = 400 \frac{\left(\frac{F_L R_c}{100}\right)^{0.42}}{\left(\frac{F_L R_c}{100}\right)^{0.42} + 27.13} + 0.1
\]
If $R_c$ is negative, then
\[
  R_a = -400 \frac{\left(\frac{-F_L R_c}{100}\right)^{0.42}}{\left(\frac{-F_L R_c}{100}\right)^{0.42} + 27.13} + 0.1
\]
and similarly for the computations of $G_a$ and $B_a$.
\end{step}

If the $\sign$ operator is used here as it is used later in step 5 of the
inverse model, the above description can be shortened.

Furthermore, the term $0.1$ is added here, but in all of the following steps in
which $R_a$ is used -- except the computation of $t$ in Step 9 --, it cancels
out algebraically.  Unfortunately, said cancellation is not always exact when
computed in floating point arithmetic. Luckily, the adverse effect of such
rounding errors is rather limited here. The results will only be distorted for
very small input values, e.g., $X=Y=Z=0$; see Table~\ref{tab:zero}. For the
sake of consistency, it is advisable to include the term $0.1$ only in the
computation of $t$ in Step 9:
\[
  R'_a = 400 \sign(R_c) \frac{{\left(\frac{F_L \abs{R_c}}{100}\right)}^{0.42}}{{\left(\frac{F_L \abs{R_c}}{100}\right)}^{0.42} + 27.13}.
\]

\begin{table}\centering
  \begin{tabularx}{\linewidth}{XXX}
  \toprule
          & with fixes & without\\
  \midrule
    $J$ & \texttt{0.0} & \texttt{3.258e-22}\\
    $C$ & \texttt{0.0} & \texttt{4.071e-24}\\
    $h$ & \texttt{0.0} & \texttt{0.0}\\
    $Q$ & \texttt{0.0} & \texttt{2.233e-10}\\
    $M$ & \texttt{0.0} & \texttt{2.943e-24}\\
    $s$ & \texttt{0.0} & \texttt{1.148e-05}\\
  \bottomrule
\end{tabularx}
\caption{CAM16 values upon input $X=Y=Z=0$ with and without the fixes in
  this article.  The exact solutions are zeros for every
  entry.}\label{tab:zero}
\end{table}

\subsection{Linear combinations, forward model}

In the forward model, four linear combinations of $R'_a$, $G'_a$, and $B'_a$
have to be formed. They can conveniently be expressed as the matrix-vector
multiplication
\[
  \begin{pmatrix}
    p'_2\\[0.5ex]
    a\\[0.5ex]
    b\\[0.5ex]
    u
  \end{pmatrix}
  \coloneqq
  \begin{pmatrix}
    2 & 1 & \tfrac{1}{20}\\[0.5ex]
    1 & -\tfrac{12}{11} & \tfrac{1}{11}\\[0.5ex]
    \tfrac{1}{9} & \tfrac{1}{9} & -\tfrac{2}{9}\\[0.5ex]
    1 & 1 & \tfrac{21}{20}
  \end{pmatrix}
  \begin{pmatrix}
    R'_a\\G'_a\\B'_a
  \end{pmatrix}
\]
which on many platforms can be computed significantly faster than four
individual dot-products.
The last variable $u$ is used in the computation of $t$ in step 9.

\subsection{Step 9, forward model}

\begin{step}[9]
  Calculate the correlates of [\dots] saturation ($s$).
  \[
    s \coloneqq 100 \sqrt{M/Q}.
  \]
\end{step}
This expression is not well-defined if $Q=0$, a value occurring if the
input values are $X=Y=Z=0$. When making use of the definition of $M$ and $Q$,
one gets to an expression for $s$ that is well-defined in all cases:
\begin{align}
  \label{eq:alpha}
  \alpha&\coloneqq t^{0.9} {(1.64-0.29^n)}^{0.73},\\
  \nonumber
  s &\coloneqq 50 \sqrt{\frac{c\alpha}{A_w + 4}}.
\end{align}

\subsection{Steps 2 and 3, inverse model}

\begin{step}[2]
Calculate $t$, $e_t$, $p_1$, $p_2$, and $p_3$.
\begin{align*}
  t &= {\left(\frac{C}{\sqrt{\frac{J}{100}} {(1.64 - 0.29^n)}^{0.73}}\right)}^\frac{1}{0.9},\\
  e_t &= \frac{1}{4} \left[\cos(h'\pi/180\degree + 2) + 3.8\right],\\
  p_1 &= \frac{50000}{13} N_c N_{cb} e_t \frac{1}{t},\\
  p_2 &= \frac{A}{N_{bb}} + 0.305,\\
  p_3 &= \frac{21}{20}.
\end{align*}
\end{step}

\begin{step}[3]
Calculate $a$ and $b$.
If $t=0$, then $a=b=0$ and go to Step 4.
In the next computations be sure transform $h$ from degrees to radians before
calculating $\sin(h)$ and $\cos(h)$: If $\abs{\sin(h)} \ge \abs{\cos(h)}$
then
\begin{align*}
  p_4 &= \frac{p_1}{\sin(h)},\\
  b &= \frac{p_2 (2+p_3) \frac{460}{1403}}{p_4 + (2+p_3) \frac{220}{1403} \frac{\cos(h)}{\sin(h)} - \frac{27}{1403} + p_3 \frac{6300}{1403}},\\
  a &= b \frac{\cos(h)}{\sin(h)}.
\end{align*}
If $\abs{\cos(h)} > \abs{\sin(h)}$ then
\begin{align*}
  p_5 &= \frac{p_1}{\cos(h)},\\
  a &= \frac{p_2 (2+p_3) \frac{460}{1403}}{%
    p_5
    + (2+p_3) \frac{220}{1403} -
    \left(\frac{27}{1403}  - p_3 \frac{6300}{1403}\right) \frac{\sin(h)}{\cos(h)}
  },\\
  b &= a \frac{\sin(h)}{\cos(h)}.
\end{align*}
\end{step}

Some of the complications in this step stem from the fact that the variable $t$
might be $0$ in the denominator of $p_1$. Likewise, the distinction of cases in $\sin(h)$ and $\cos(h)$ is necessary
to avoid division by $0$ in $a$ and $b$.

It turns out that both of these problems can be avoided quite elegantly.
Consider, in the case $\abs{\sin(h)} \ge \abs{\cos(h)}$:
\begin{align*}
  p'_1 &\coloneqq \frac{50000}{13} N_c N_{cb} e_t,\\
  b &= \frac{p_2 (2+p_3) \frac{460}{1403}}{\frac{p'_1}{t\sin(h)} + (2+p_3) \frac{220}{1403} \frac{\cos(h)}{\sin(h)} - \frac{27}{1403} + p_3 \frac{6300}{1403}}\\
   &= \frac{t \sin(h) p_2 (2+p_3) \frac{460}{1403}}{p'_1 + t (2+p_3) \frac{220}{1403} \cos(h) + t \sin(h) \frac{6588}{1403}}\\
   &= \frac{23 t \sin(h) p_2}{23 p'_1 + 11 t \cos(h) + 108 t \sin(h)},
\end{align*}
and
\[
  a = \frac{23 t \cos(h) p_2}{23 p'_1 + 11 t \cos(h) + 108 t \sin(h)}.
\]
Conveniently, the exact same expressions are retrieved in the case
$\abs{\cos(h)} > \abs{\sin(h)}$. These expressions are always well-defined since
\begin{multline*}
  23 p'_1 + 11 t \cos(h) + 108 t \sin(h)\\
  = \frac{23 p'_1 p_2}{R'_a + G'_a + \tfrac{21}{20}B'_a + 0.305}
  > 0.
\end{multline*}

In the algorithm, the value of $t$ can be retrieved via
$\alpha$~\eqref{eq:alpha} from the input variables. Indeed, if the saturation
correlate $s$ is given, one has
\[
  \alpha \coloneqq {\left(\frac{s}{50}\right)}^2 \frac{A_w+4}{c};
\]
if $M$ is given, one can compute $C\coloneqq M / F_L^{0.25}$ and then
\[
\alpha\coloneqq\begin{dcases*}
  0 &if $J=0$,\\
  \frac{C}{\sqrt{J/100}}&otherwise.
\end{dcases*}
\]
It is mildly unfortunate that one has to introduce a case distinction for
$J=0$ here, but this is an operation that can still be performed at reasonable
efficiency.

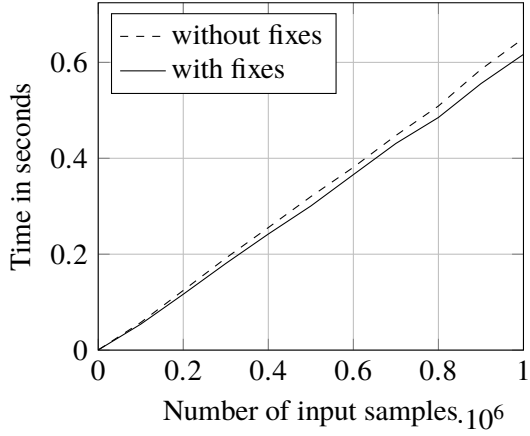
\begin{figure}
% This file was created by matplotlib2tikz v0.6.15.
\begin{tikzpicture}

\begin{axis}[
width=\linewidth,
xlabel={Number of input samples},
ylabel={Time in seconds},
xmin=0, xmax=1000000,
ymin=0.0, ymax=0.724347234702873,
xmajorgrids,
ymajorgrids,
legend cell align={left},
legend entries={{without fixes},{with fixes}},
legend style={at={(0.03,0.97)}, anchor=north west}
]
\addplot [dashed, black]
table {%
0 0.000386412008083425
100000 0.0575982900045346
200000 0.124436895988765
300000 0.191784891998395
400000 0.255525363012566
500000 0.320106088009197
600000 0.381167707993882
700000 0.447597951002535
800000 0.50900621600158
900000 0.585287615002017
1000000 0.649983358001919
};
\addplot [black]
table {%
0 0.000280210006167181
100000 0.0540978200006066
200000 0.116504425997846
300000 0.1806468910072
400000 0.242063231999055
500000 0.300529030006146
600000 0.365901857003337
700000 0.430936441989616
800000 0.485103309008991
900000 0.555676405012491
1000000 0.616663155989954
};
\end{axis}

\end{tikzpicture}
  \caption{Performance comparison of the conversion from CAM16 to XYZ (with
  $J$, $C$, and $h$), implemented in colorio~\cite{colorio}. The suggested
  improvements in the inverse model lead to a speed-up of about 5\%.}
\end{figure}

\appendix
\section{Full model\label{sec:full}}

For the convenience of the reader, both forward and inverse steps of the
improved CAM16 algorithm are given here. The wording is taken from~\cite{cam16}
where applicable.
The steps that differ from the original model are marked with an asterisk~(*).

As an abbreviation, the bold letter $\rgb$ is used whenever the equation applies
to $R$, $G$, and $B$ alike.

\paragraph{Illuminants, viewing surrounds set up and background
parameters}
(See the note at the end of Part 2 of Appendix B of~\cite{cam16} for determining
all parameters.)

\begin{itemize}
  \item Adopted white in test illuminant: $X_w$, $Y_w$, $Z_w$
  \item Background in test conditions: $Y_b$
  \item Reference white in reference illuminant:
    $X_{wr}=Y_{wr} = Z_{wr}=100$, fixed in the model
  \item Luminance of test adapting field (\si{\candela\per\meter\squared}): $L_A$.
$L_A$ is computed using
    \[
      L_A = \frac{E_W}{\pi} \frac{Y_b}{Y_W} = \frac{L_W Y_b}{Y_W},
    \]
  where $E_W =\pi L_W$ is the illuminance of reference white in \si{\lux};
    $L_W$ is the luminance of reference white in
    \si{\candela\per\meter\squared}; $Y_b$ is the luminance factor of the
    background; and $Y_w$ is the luminance factor of the reference white.

  \item Surround parameters are given in Table~\ref{tab:surround}:
To determine the surround conditions see the note at the
    end of Part 1 of Appendix A of~\cite{cam16}.
\item $N_c$ and $F$ are modelled as a function of $c$, and their values
  can be linearly interpolated, using the data from~\ref{tab:surround}.
\end{itemize}

Let $M_{16}$ be given by
\[
  M_{16} \coloneqq \begin{pmatrix}
    0.401288  & 0.650173 & -0.051461\\
    -0.250268 & 1.204414 & 0.045854\\
    -0.002079 & 0.048952 & 0.953127
  \end{pmatrix}.
\]

\subsection{Forward model}

\begin{step}[0*]
Calculate all values/parameters which are independent
of the input sample.
\begin{align*}
  &\begin{pmatrix}R_w\\G_w\\B_w\end{pmatrix}
    = M_{16}
  \begin{pmatrix}X_w\\Y_w\\Z_w\end{pmatrix},\\
  &D = F \left[1 - \tfrac{1}{3.6} \exp\left(\tfrac{-L_a-42}{92}\right)\right].
\end{align*}
If $D$ is greater than one or less than zero, set it to one or zero,
respectively.
\begin{align*}
  &D_{\rgb} = D\frac{Y_W}{\rgb_W} -1 + D,\\
  &k = \frac{1}{5L_A + 1},\\
  &F_L = k^4 L_A + 0.1 {(1-k^4)}^2 {(5L_A)}^{1/3},\\
  &n = \frac{Y_b}{Y_W},\\
  &z = 1.48 + \sqrt{n},\\
  &N_{bb} = \frac{0.725}{n^{0.2}},\\
  &N_{cb} = N_{bb},\\
  &\rgb_{wc} = D_{\rgb} \rgb_w,\\
  &\rgb_{aw} = 400
  \frac
  {{\left(\frac{F_L \rgb_{wc}}{100}\right)}^{0.42}}
  {{\left(\frac{F_L \rgb_{wc}}{100}\right)}^{0.42} + 27.13},\\
  &A_w = \left(2R_{aw} + G_{aw} + \tfrac{1}{20} B_{aw}\right) \cdot N_{bb}.
\end{align*}
\end{step}

\begin{table}\centering
  \begin{tabularx}{\linewidth}{XXXX}
  \toprule
          & $F$ & $c$   & $N_c$\\
  \midrule
  Average & 1.0 & 0.69  & 1.0\\
  Dim     & 0.9 & 0.59  & 0.9\\
  Dark    & 0.8 & 0.525 & 0.8\\
  \bottomrule
\end{tabularx}
  \caption{Surround parameters.}\label{tab:surround}
\end{table}

\begin{table}\centering
  \begin{tabularx}{\linewidth}{XXXXXX}
  \toprule
        & Red   & Yellow & Green & Blue   & Red\\
  \midrule
  $i$   & 1     & 2     & 3      & 4      & 5\\
  $h_i$ & 20.14 & 90.00 & 164.25 & 237.53 & 380.14\\
  $e_i$ & 0.8   & 0.7   & 1.0    & 1.2    & 0.8\\
  $H_i$ & 0.0   & 100.0 & 200.0  & 300.0  & 400.0\\
  \bottomrule
\end{tabularx}
  \caption{Unique hue data for calculation of hue quadrature.}\label{table:hue}
\end{table}

\begin{step}[1]
Calculate `cone' responses.
\[
\begin{pmatrix}R\\G\\B\end{pmatrix}
= M_{16} \begin{pmatrix}X\\Y\\Z\end{pmatrix}
\]
\end{step}

\begin{step}[2]
Complete the color adaptation of the illuminant in
the corresponding cone response space (considering various
luminance levels and surround conditions included in $D$, and
hence in $D_R$, $D_G$, and $D_B$).
\[
  \rgb_c = D_{\rgb} \cdot \rgb
\]
\end{step}

\begin{step}[3*]
Calculate the modified postadaptation cone response
(resulting in dynamic range compression).
\[
  \rgb'_a = 400 \sign(\rgb_c)
    \frac
    {{\left(\frac{F_L \abs{\rgb_c}}{100}\right)}^{0.42}}
    {{\left(\frac{F_L \abs{\rgb_c}}{100}\right)}^{0.42} + 27.13}.
\]
\end{step}

\begin{step}[4*]
Calculate Redness--Greenness ($a$), Yellowness--Blueness ($b$) components,
  hue angle ($h$), and auxiliary variables ($p'_2$, $u$).
\begin{align*}
  \begin{pmatrix}
    p'_2\\[0.5ex]
    a\\[0.5ex]
    b\\[0.5ex]
    u
  \end{pmatrix}
  &\coloneqq
  \begin{pmatrix}
    2 & 1 & \tfrac{1}{20}\\[0.5ex]
    1 & -\tfrac{12}{11} & \tfrac{1}{11}\\[0.5ex]
    \tfrac{1}{9} & \tfrac{1}{9} & -\tfrac{2}{9}\\[0.5ex]
    1 & 1 & \tfrac{21}{20}
  \end{pmatrix}
  \begin{pmatrix}
    R'_a\\G'_a\\B'_a
  \end{pmatrix},\\
  % a&\coloneqq R'_a - \tfrac{12}{11} G'_a + \tfrac{1}{11} B'_a\\
  % b&\coloneqq \tfrac{1}{9} R'_a + \tfrac{1}{9} G'_a - \tfrac{2}{9} B'_a\\
  h&\coloneqq \arctan(b/a).
\end{align*}
(Make sure that $h$ is between $0\degree$ and $360\degree$.)
\end{step}

\begin{step}[5]
Calculate eccentricity [$e_t$, hue quadrature composition
($H$) and hue composition ($H_c$)].

Using the following unique hue data in table~\ref{table:hue}, set
$h'= h + 360\degree$ if $h < h_1$, otherwise $h'=h$.
Choose a proper $i\in\{1,2,3,4\}$ so that $h_i\le h' < h_{i+1}$.
Calculate
\[
  e_t = \tfrac{1}{4}
  \left[
    \cos(h'\pi/180\degree + 2) + 3.8
  \right]
\]
which is close to, but not exactly the same as, the eccentricity factor given
in table~\ref{table:hue}.

Hue quadrature is computed using the formula
\[
  H = H_i + \frac{100 e_{i+1} (h'-h_i)}{e_{i+1}(h'-h_i) + e_i (h_{i+1}-h')}
\]
and hue composition $H_c$ is computed according to $H$.  If $i=3$ and $H =
241.2116$ for example, then $H$ is between $H_3$ and $H_4$ (see
table~\ref{table:hue} above). Compute $P_L=H_4-H = 58.7884$; $P_R = H – H_3 =
41.2116$ and round $P_L$ and $P_R$ values to integers $59$ and $41$. Thus,
according to table~\ref{table:hue}, this sample is considered as having 59\%
of green and 41\% of blue, which is the $H$c and can be reported as 59G41B or
41B59G.
\end{step}

\begin{step}[6*]
Calculate the achromatic response
\[
  A\coloneqq p'_2 \cdot N_{bb}.
  \]
\end{step}

\begin{step}[7]
Calculate the correlate of lightness
\[
  J \coloneqq 100 {(A / A_w)}^{cz}.
\]
\end{step}

\begin{step}[8]
  Calculate the correlate of brightness
  \[
    Q \coloneqq \frac{4}{c} \sqrt{\frac{J}{100}} (A_w+4) F_L^{0.25}.
    \]
\end{step}

\begin{step}[9*]
Calculate the correlates of chroma ($C$), colorfulness ($M$), and saturation
  ($s$).
\begin{align*}
  t&\coloneqq \frac{50000/13 N_c N_{cb} e_t \sqrt{a^2 + b^2}}{u + 0.305},\\
  \alpha&\coloneqq t^{0.9} {(1.64 - 0.29^n)}^{0.73},\\
  C&\coloneqq \alpha \sqrt{\frac{J}{100}},\\
  M&\coloneqq C\cdot F_L^{0.25},\\
  s &\coloneqq 50 \sqrt{\frac{\alpha c}{A_w + 4}}.
\end{align*}
\end{step}

\subsection{Inverse model}

\begin{step}[1]
  Obtain $J$, $t$, and $h$ from $H$, $Q$, $C$, $M$, $s$.

  The input data can be different combinations of perceived correlates, that
  is, $J$ or $Q$; $C$, $M$, or $s$; and $H$ or $h$. Hence, the following
  sub-steps are needed to convert the input parameters to the parameters $J$,
  $t$, and $h$.
\end{step}

\begin{step}[1--1]
Compute $J$ from $Q$ (if input is $Q$)
\[
  J\coloneqq 6.25 \frac{cQ}{(A_w+4) F_L^{0.25}}.
\]
\end{step}

\begin{step}[1--2*]
Calculate $t$ from $C$, $M$, or $s$.
\begin{itemize}
  \item If input is $C$ or $M$:
    \begin{align*}
      C &\coloneqq M / F_L^{0.25} \:\text{if input is $M$}\\
      \alpha &\coloneqq \begin{dcases*}
          0 &if $J=0$,\\
          \frac{C}{\sqrt{J/100}}& otherwise.
      \end{dcases*}
    \end{align*}
  \item If input is $s$:
    \[
    \alpha \coloneqq {\left(\frac{s}{50}\right)}^2 \frac{A_w+4}{c}
    \]
\end{itemize}
Compute $t$ from $\alpha$:
\[
  t \coloneqq {\left(\frac{\alpha}{{(1.64 - 0.29^n)}^{0.73}}\right)}^{1/0.9}
\]
\end{step}

\begin{step}[1--3]
Calculate $h$ from $H$ (if input is $H$).
The correlate of hue ($h$) can be computed by using data in
table~\ref{table:hue} in the forward model.
Choose a proper $i\in\{1,2,3,4\}$ such that
$H_i \le H < H_{i+1}$. Then
\[
  h' = \frac{(H-H_i)(e_{i+1}h_i - e_i h_{i+1}) - 100 h_i e_{i+1}}{(H-H_i)(e_{i+1}-e_i) - 100 e_{i+1}}.
\]
Set $h = h' - 360\degree$ if $h' > 360\degree$, and $h=h'$ otherwise.
\end{step}

\begin{step}[2*]
Calculate $e_t$, $A$, $p'_1$, and $p'_2$
\begin{align*}
  e_t &= \tfrac{1}{4} (\cos(h\pi/180\degree + 2) + 3.8),\\
  A &= A_w  {(J/100)}^{1/(cz)},\\
  p'_1 &= e_t \tfrac{50000}{13} N_c N_{cb},\\
  p'_2 &= A / N_{bb}.
\end{align*}
\end{step}

\begin{step}[3*]
Calculate $a$ and $b$
  \begin{align*}
    \gamma &\coloneqq \frac{23 (p'_2+0.305) t}{23 p'_1 + 11 t \cos(h) + 108 t \sin(h)},\\
    a &\coloneqq \gamma \cos(h),\\
    b &\coloneqq \gamma \sin(h).
  \end{align*}
\end{step}

\begin{step}[4]
  Calculate $R'_a$, $G'_a$, and $B'_a$.
  \[
  \begin{pmatrix}
    R'_a\\G'_a\\B'_a
  \end{pmatrix}
  =
  \frac{1}{1403}
  \begin{pmatrix}
    460 & 451 & 288\\
    460 & -891 & -261\\
    460 & -220 & -6300
  \end{pmatrix}
  \begin{pmatrix}
    p'_2\\a\\b
  \end{pmatrix}.
  \]%
\end{step}

\begin{step}[5*]
Calculate $R_c$, $G_c$, and $B_c$,
  \[
  \rgb_c = \sign(\rgb'_a)
  \frac{100}{F_L} {\left(
    \frac{27.13 \abs{\rgb'_a}}{400 - \abs{\rgb'_a}}
    \right)}^{1/0.42}.
  \]
\end{step}

\begin{step}[6]
Calculate $R$, $G$, and $B$ from $R_c$, $G_c$, and $B_c$.
\[
  \rgb = \rgb_c / D_{\rgb}.
\]
\end{step}

\begin{step}[7]
Calculate $X$, $Y$, and $Z$. (For the coefficients of the inverse matrix, see
the note at the end of the appendix B of~\cite{cam16}.)
\[
\begin{pmatrix}X\\Y\\Z\end{pmatrix}
  = M_{16}^{-1}
\begin{pmatrix}R\\G\\B\end{pmatrix}.
\]
\end{step}

% \printbibliography{}
\bibliography{bib}{}
\bibliographystyle{plain}

\end{document}